Annealing-Dependent Magnetic Depth Profile in $Ga_{1-x}Mn_xAs$

B. J. Kirby

Department of Physics and Astronomy, University of Missouri, Columbia, Missouri 65211

J. A. Borchers

NIST Center for Neutron Research, National Institute of Standards and Technology,

Gaithersburg, Maryland 20899

J. J. Rhyne

Lujan Neutron Scattering Center, Los Alamos National Laboratory, Los Alamos, New Mexico

87545

and Department of Physics and Astronomy, University of Missouri, Columbia, Missouri 65211

S. G. E. te Velthuis and A. Hoffmann

Materials Science Division, Argonne National Laboratory, Argonne, Illinois 60439

K. V. O'Donovan

NIST Center for Neutron Research, National Institute of Standards and Technology,

Gaithersburg, Maryland 20899

and Department of Materials Science and Engineering, University of Maryland, College Park,

Maryland 20742

T. Wojtowicz

Department of Physics, University of Notre Dame, Notre Dame, Indiana 46556

and Institute of Physics of the Polish Academy of Sciences, 02-688 Warsaw, Poland

X. Liu, W. L. Lim, and J. K. Furdyna

Department of Physics, University of Notre Dame, Notre Dame, Indiana 46556

We have studied the depth-dependent magnetic and structural properties of as-grown and optimally annealed $Ga_{1-x}Mn_xAs$ films using polarized neutron reflectometry. In addition to increasing the total magnetization, the annealing process was observed to produce a significantly more homogeneous distribution of the magnetization. This difference in the films is attributed to the redistribution of Mn at interstitial sites during the annealing process. Also, we have seen evidence of significant magnetization depletion at the surface of both as-grown and annealed films.

Recently, there has been a great deal of interest in the development of high Curie temperature ($T_C$) ferromagnetic semiconductors for use in spintronics applications. $Ga_{1-x}Mn_xAs$ is a possible candidate for such applications,[1,2] with $T_C$ reaching 150 K in some cases.[3] The ferromagnetic behavior in this material originates from coupling between spin 5/2 $Mn^{2+}$ ions substituting for Ga.[4] These substitutional Mn ions ($Mn_{Ga}$) act as acceptors, generating holes that mediate the ferromagnetic exchange. However, $Mn_{Ga}$ are known to be partially compensated by other impurities, such as As at Ga sites ($As_{Ga}$),[5,6] and Mn at interstitial sites ($Mn_I$),[7,8,9] which are double donors.

Magnetization measurements of $Ga_{1-x}Mn_xAs$ typically show the magnetic moment per Mn atom to be less than the value of 4 $\mu_B$ that would be expected for spin 5/2 divalent Mn, indicating that not all of the Mn atoms participate in the ferromagnetic exchange.[10] This is at least partially due to $Mn_I$ aligning antiferromagnetically with $Mn_{Ga}$, effectively canceling their moments.[11] It has been well established that low temperature post-growth annealing of $Ga_{1-x}Mn_xAs$ films can serve to significantly raise $T_C$,[12] and increase the magnetization.[9,13] Yu et al.[8,9] present evidence to suggest that this phenomenon is in large part due to the redistribution of $Mn_I$ during annealing.

In this paper, we present a study of the magnetic and structural depth profiles of as-grown and optimally annealed $Ga_{1-x}Mn_xAs$ thin films grown by molecular-beam epitaxy (MBE). A $Ga_{1-x}Mn_xAs$ film was prepared by first depositing a 300 nm GaAs buffer layer on a (001) GaAs substrate at a temperature of 580 °C, then cooling the substrate to 210 °C and adding another 3 nm GaAs buffer layer, before depositing 115 ± 10 nm of $Ga_{1-x}Mn_xAs$ at 210 °C. Using x-ray diffraction, the Mn concentration in the film was estimated to be $x = 0.073 ± 0.01$.[14] This film was then cleaved into two pieces. One piece was annealed in $N_2$ for 1 hour, at a temperature of 280 °C, while the other piece was left as-grown. Resistivity measurements indicated that annealing increased $T_C$ from 60 K to 125 K.

The as-grown and annealed films were then examined by polarized neutron reflectometry (PNR) using the NG-1 Reflectometer at the NIST Center for Neutron Research. In our experiments, a magnetic field $H$ was applied in the plane of the film. Neutrons were polarized using Fe/Si supermirrors in combination with Al-coil spin flippers to have their spin polarization oriented either parallel or anti-parallel to $H$, and were specularly reflected from the film. The reflectivity was measured as a function of wavevector transfer $Q$ for both spin-flip (neutrons incident and reflected with opposite polarizations), and non spin-flip (neutrons incident and reflected with the same polarization) scattering cross-sections.

By exploiting the wave nature and magnetic moment of the neutron, PNR provides the unique ability to establish depth profiles of the structure, and of the vector magnetization in thin film samples.[15,16] Specifically, the reflectivity can be fit[17,18] using a depth dependent scattering length density (SLD) profile $\rho(z)$ (where $z$ is the film depth) with nuclear and magnetic components,

$$\rho(z) = \rho_{nuc}(z) \pm \rho_{mag}(z), \tag{1}$$

$$\rho_{nuc}(z) = \sum_i N_i(z) b_i, \tag{2}$$

$$\rho_{mag}(z) = C \sum_i N_i(z)\, \mu_i, \qquad (3)$$

where the summation is over each type of atom in the system, $N$ is the in-plane average of the number density, $b$ is the nuclear scattering length, and $\mu$ is the magnetic moment in Bohr magnetons. The constant $C = 2.69\ \mathrm{fm}/\mu_B$. The various types of Mn in this system each have the same value of $b$, but different values of $\mu$ – therefore, the above summations include the individual counting of each separate type of Mn. The sign before $\rho_{mag}$ in Eq. (1) depends on the orientation of the magnetization relative to the neutron polarization.

For our films, the scattering from the nuclear structure was significantly stronger than the magnetic scattering due to the low Mn concentration. To maximize the magnetic scattering, all PNR measurements shown here were taken at a temperature of 13 K, and in an applied in-plane magnetic field of 1 kOe after zero-field cooling the films.

The two non spin-flip (NSF) reflectivities $R_{++}(Q)$ and $R_{--}(Q)$ for both the as-grown and annealed films are shown in Fig. 1, along with fits to the data generated from the corresponding SLD model. To better accentuate their features, the reflectivities and fits have been multiplied by $Q^4$, and are shown on a logarithmic scale. The splitting between the $R_{++}$ and the $R_{--}$ reflectivities originates from the component of each film's magnetization parallel to $\boldsymbol{H}$[15], with the magnitude of the splitting being indicative of the magnetization at a particular length scale. While the $R_{--}$ reflectivity shows somewhat similar oscillations for both films, the two films have very different $R_{++}$ reflectivities. For the as-grown film, the $R_{++}$ reflectivity shows pronounced oscillations that are slightly phase shifted with respect to its companion $R_{--}$ oscillations. By comparison, the annealed film's $R_{++}$ reflectivity is very smooth, without well defined oscillations. Because of this behavior, fits to the data reveal differences in the depth-dependent magnetic properties of the two films that extend beyond differences in their net magnetization.

The spin-flip (SF) reflectivities were measured to be at background levels for both films, and are not shown in Fig. 1. The presence of SF scattering would have indicated a component of the film's magnetization perpendicular to $H$.[15] Therefore, its absence means that we do not observe evidence of coherent moment canting at these field and temperature conditions.

Since our systems appear to be magnetically saturated, it is useful to recast the reflectivities in terms of spin asymmetry,

$$SA(Q) = \frac{R_{++}(Q) - R_{--}(Q)}{R_{++}(Q) + R_{--}(Q)}. \tag{4}$$

The spin asymmetry accentuates the scattering from the component of the magnetization parallel to $H$, and provides an intuitive way of gauging the magnetization at different length scales.

The measured spin asymmetries and those from the fits to the reflectivity for the as-grown and annealed films are shown in Fig. 2. The peak amplitudes of the spin asymmetry at low $Q$ are largely determined by the magnitude of the net magnetization of the film, and show the expected increase in magnetization upon annealing. Additionally, the spin asymmetry for the annealed film displays oscillations that are better defined than those for the as-grown film. Since a smearing of the oscillations can be indicative of magnetic roughness, these data suggest that the annealed film possesses a more uniform magnetization than the as-grown film.

The SLD models used to successfully fit the data are shown in Fig. 3 with $\rho_{nuc}$ and $\rho_{mag}$ plotted as functions of film depth. The depth resolution for features in the models is approximately 5 Å. Since $\rho_{mag}$ is directly proportional to the magnetization $M$ of the film, the magnetization scale is also shown. Integrating $M$ over $z$, and dividing by the total film thickness gives the average film magnetization, $M_{avg}$ (inset in Fig. 3) that can be compared to net values. The SLD models show $M_{avg}$ = 17 emu cm$^{-3}$ for the as-grown film (approximately 1.1 $\mu_B$ per Mn$_{Ga}$), and $M_{avg}$ = 48 emu cm$^{-3}$ for the annealed film (approximately 3.3 $\mu_B$ per Mn$_{Ga}$). This shows the expected result that more of the Mn ions are participating in the ferromagnetic exchange after annealing.

However, what is striking about these results is the difference in *depth distribution* of the magnetization between the two films. It is immediately noticeable that the SLD profiles of the two films are different. These differences can be interpreted in part by considering the unique signature that Mn leaves on both the nuclear and the magnetic SLD profiles. Mn (at any lattice site or other random location) should be the only atom in this system with a *negative nuclear scattering length*. This means that a decrease (increase) in $\rho_{nuc}$ generally implies an increased (decreased) concentration of Mn. Additionally, Mn$_{Ga}$ should be the only atom in this system significantly contributing to the ferromagnetic exchange. This means that an increase (decrease) in $\rho_{mag}$ generally implies an increased (decreased) concentration of Mn$_{Ga}$ *uncompensated* by Mn$_I$.

It should be pointed out that recent x-ray magnetic circular dichroism measurements have revealed the presence of induced magnetic moments on Ga and As atoms in $Ga_{1-x}Mn_xAs$.[19] However, these induced moments are thought to be very small compared to the Mn moment,[20,21] and are unlikely to be responsible for depth-dependent changes in film magnetization of the scale reported in this paper. Additionally, it is unlikely that changes in $As_{Ga}$ distribution contribute to annealing-dependent differences, as it is a relatively stable defect at our annealing temperatures.[22] Therefore, most of the non-uniformity in the profiles can be attributed to variations in Mn concentration and/or site occupation.

Starting at the substrate of the as-grown film, the top panel of Fig. 3 shows that $\rho_{nuc}$ decreases as $\rho_{mag}$ increases, indicating an increase in Mn concentration at the substrate interface. Above that interface, $\rho_{mag}$ gradually climbs, peaking at about 100 Å from the free surface. Over that same region, $\rho_{nuc}$ is very uniform, indicating that the total Mn concentration is nearly constant as the free surface is approached. Therefore, comparison of the $\rho_{nuc}$ and $\rho_{mag}$ profiles suggests that the concentration of *uncompensated* $Mn_{Ga}$ progressively increases. This could indicate that during the growth process, formation of $Mn_{Ga}$ is more favorable just below the free surface. At 40 Å from the free surface $\rho_{mag}$ rapidly drops to zero, while $\rho_{nuc}$ also drops. This suggests that there is a slightly increased total Mn concentration at the free surface, but that virtually none of the free surface Mn is contributing to the ferromagnetic exchange. However, there is some added uncertainty surrounding this small increase in surface Mn, as the free surface roughness and the free surface value of $\rho_{nuc}$ are somewhat tenuous features of this model.

In stark comparison with the as-grown film, the annealed film's magnetic SLD profile is relatively constant for most of its thickness. However, it too has important features. Again, starting at the substrate,[23] the bottom panel of Fig. 3 shows a buildup of Mn concentration, and a gradually increasing magnetization that does not level off until 900 Å from the free surface. There is also a slight increase in $\rho_{mag}$ and $\rho_{nuc}$ over a 500 Å region, starting at 800 Å from the free surface. At 90 Å from the free surface, $\rho_{mag}$ drastically drops as $\rho_{nuc}$ drastically rises – all the way to the value of the substrate. One simple interpretation of this is that the surface layer has little to no Mn present. However, recently reported measurements[24] provide evidence of *increased* Mn concentration at the free surface of annealed $Ga_{1-x}Mn_xAs$ films, which is attributed to out-diffusion of $Mn_I$. Additionally, the models in Fig. 3 suggest that the annealed film is slightly thicker than the as-grown film. This leads to consideration of a different interpretation, that the free surface features of the annealed film indicate the presence of a compound with a SLD profile *very similar* to that of GaAs, such as antiferromagnetic $\theta$-MnN.[25] Therefore, it is possible that MnN or a related compound may have formed at the surface during annealing in nitrogen. Since PNR cannot distinguish between these two possible interpretations, investigations using other methods will be required to fully resolve this issue.

PNR data for a second set of as-grown and annealed films measured using both the POSY I Reflectometer at the Argonne Intense Pulsed Neutron Source and NG-1 at NIST, are similar to those shown in Fig. 2. SLD models used to fit those measurements were comparable to the ones shown in Fig. 3. Both the annealed and as-grown films again exhibited a depletion of magnetization at the surface, while only the as-grown film featured a positive gradient of magnetization as the surface was approached. The reproducibility of these effects suggests that the annealing dependence of the magnetization distribution, as well as the surface magnetization depletion could be *general* properties of MBE-grown $Ga_{1-x}Mn_xAs$ with $x \approx 0.07$.

To summarize, we have demonstrated that polarized neutron reflectometry, typically applied to the characterization of concentrated magnetic systems, can also provide detailed information about the spatial distribution of magnetic ions in very dilute ferromagnetics, such as $Ga_{1-x}Mn_xAs$ with $x$ as low as 0.07. We have also provided independent evidence that low temperature post-growth annealing, in addition to increasing $T_C$, also increases the total magnetization in $Ga_{1-x}Mn_xAs$, as has been previously reported on the basis of SQUID studies.[8,10] Our studies additionally show, for the first time, that annealing produces a more homogeneous distribution of the magnetization as a function of depth. This result strongly corroborates the concept[8,9] that the annealing process redistributes $Mn_I$, possibly to the surface, where it doesn't cancel the magnetic moment of existing $Mn_{Ga}$. Additionally, for both the as-grown, and the annealed films, we find evidence for drastically reduced magnetization at the free surface.

The contribution to this work from Missouri and Notre Dame was supported by NSF Grant DMR-013819. Work at Argonne was supported by US DOE, Office of Science contract #W-31-109-ENG-38. Special thanks go to Paul Kienzle of NIST for development of reflectivity fitting software, and to Chuck Majkrzak of NIST and Kevin Edmonds of the University of Nottingham for helpful discussions.

Figure Captions

FIG. 1. Measured NSF reflectivities for each film, along with fits to the data from the corresponding SLD model. Polarization efficiency and footprint corrections have been applied to the data. The data and fits have been multiplied by $Q^4$, and are shown on a logarithmic scale in order to highlight their features. The reflectivity of the as-grown film has been offset by an order of magnitude to allow for comparison.

FIG. 2. The measured spin asymmetries for each film, along with the fits from the corresponding SLD model.

FIG. 3. Scattering length density models for each film. The magnetization is proportional to the magnetic component, and is shown on the right.

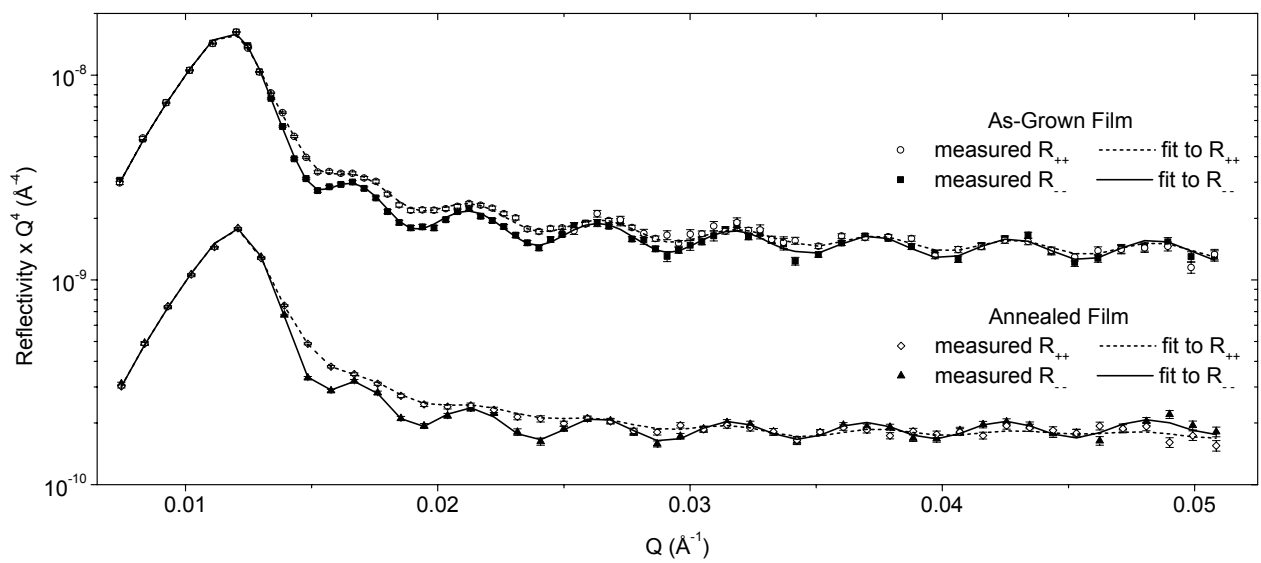

Figure 1, B. J. Kirby. This figure should be printed as two columns wide.

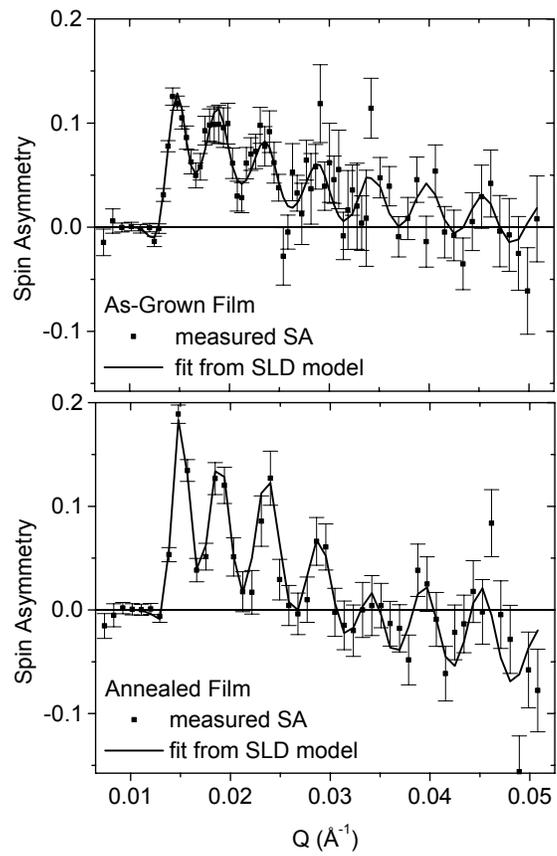

Figure 2, B. J. Kirby.

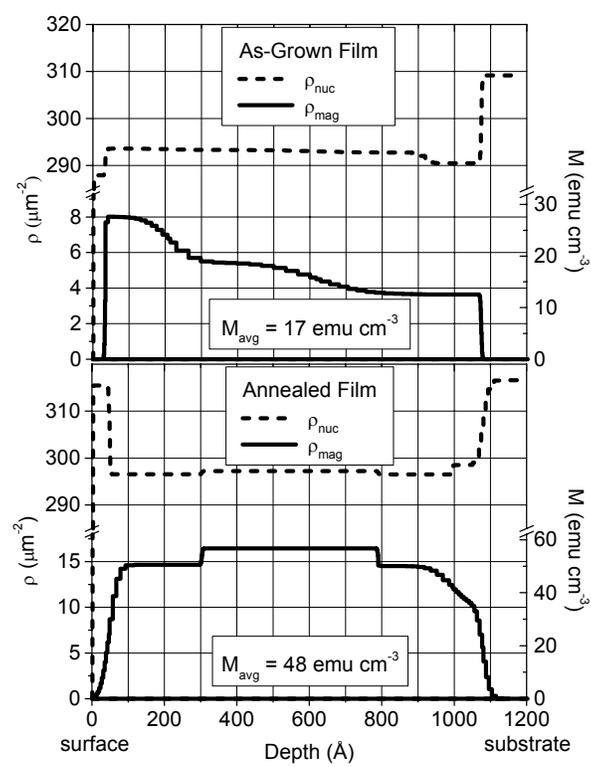

Figure 3, B. J. Kirby.